\documentclass[conference]{IEEEtran}
\IEEEoverridecommandlockouts

\usepackage{cite}
\usepackage{amsmath,amssymb,amsfonts}
\usepackage{algorithmic}
\usepackage{graphicx}
\usepackage{textcomp}
\usepackage{xcolor}
\def\BibTeX{{\rm B\kern-.05em{\sc i\kern-.025em b}\kern-.08em T\kern-.1667em\lower.7ex\hbox{E}\kern-.125emX}}

\begin{document}

\title{Terahertz Spatial Wireless Channel Modeling with Radio Radiance Field \\
}

\author{\IEEEauthorblockN{John Song\IEEEauthorrefmark{1}, Lihao Zhang\IEEEauthorrefmark{1}, Feng Ye\IEEEauthorrefmark{2}, and 
Haijian Sun\IEEEauthorrefmark{1}}
\IEEEauthorblockA{\IEEEauthorrefmark{1}School of Electrical and Computer Engineering, University of Georgia, Athens, GA, USA 
\\
\IEEEauthorrefmark{2}Department of Electrical and Computer Engineering, University of Wisconsin-Madison, Madison, WI, USA \\
Emails: \IEEEauthorrefmark{1}\{john.song, lihao.zhang, hsun\}@uga.edu, \IEEEauthorrefmark{2}feng.ye@wisc.edu}}

\maketitle

\begin{abstract}
Terahertz (THz) communication is a key enabler for 6G systems, offering ultra-wide bandwidth and unprecedented data rates. However, THz signal propagation differs significantly from lower-frequency bands due to severe free space path loss, minimal diffraction and specular reflection, and prominent scattering, making conventional channel modeling and pilot-based estimation approaches inefficient. In this work, we investigate the feasibility of applying  radio radiance field (RRF) framework to the THz band. This method reconstructs a continuous RRF using visual-based geometry and sparse THz RF measurements, enabling efficient spatial channel state information (Spatial-CSI) modeling without dense sampling. We first build a fine simulated THz scenario, then we reconstruct the RRF and evaluate the performance in terms of both reconstruction quality and effectiveness in THz communication, showing that the reconstructed RRF captures key propagation paths with sparse training samples. Our findings demonstrate that RRF modeling remains effective in the THz regime and provides a promising direction for scalable, low-cost spatial channel reconstruction in future 6G networks.
\end{abstract}


\section{Introduction}
Wireless communication technologies have rapidly evolved in the past decades, driven by the ever-increasing demand for higher data rates, lower latency, and improved efficiency. As wireless data traffic continues to surge, next-generation systems are pushing toward Terabit-per-second (Tb/s) wireless links \cite{6882305, 9541155}. Terahertz (THz) communication, operating in the broad spectrum between 0.1 and 10 THz, has emerged as a promising solution to meet this demand. THz systems offer significantly wider bandwidths and reduced interference due to their high directivity and short transmission ranges \cite{10.1109/MWC.2024.10438977}. However, the physical characteristics of THz waves introduce significant challenges to wireless communication systems.

First, due to the extremely short wavelengths, THz signals experience severe free-space path loss compared to lower frequency bands under the traditional single-input single-output (SISO) configuration. Fortunately, this limitation can be compensated by densely packing a large number of tiny THz antenna elements (on the order of hundreds) within a small physical area. As a result, high antenna array gains at both the transmitter (Tx) and receiver (Rx) become feasible with beamforming, making such gigantic multiple-input multiple-output (MIMO) architectures a defining feature of THz communication systems.
Second, the propagation behavior of THz waves differs significantly from lower frequency bands such as sub-6 GHz or frequency range 3 (FR3, 7-24 GHz). In the lower bands, diffraction and specular reflection are the dominant mechanisms that ensure coverage in non-line-of-sight (NLoS) regions. In contrast, these two kinds of interactions are rare at THz frequencies. Instead, diffuse reflections (scattering) become more pronounced and can be partially leveraged for the NLoS region coverage.

Nonetheless, both the gigantic MIMO configuration and the reliance on scattering NLoS paths impose stringent requirements on the accuracy of spatial channel knowledge. Traditional pilot-based channel estimation methods—using full digital beamformer, sending pilots, decoding the full channel matrix, and computing beamforming weights—become increasingly impractical to scale at THz due to the gigantic size of the antenna arrays. Although sparsity-aware approaches and hybrid beamforming have been proposed to simplify this process, they are essentially capturing the few dominant propagation paths in the environment. And the goal of beamforming in such scenarios is to adaptively steer array pattern lobes towards these dominant paths to the target users and suppress the interference along the paths to other non-target users. Given the increasing pilot overhead, the cost and complexity of operating full-digital or hybrid beamformers, and the computational burden of real-time beamforming optimization, there is a growing interest in implementing the spatial beamforming configuration, which aims to identify strong propagation paths and perform beamforming accordingly. 

For this purpose, traditional spatial channel modeling approaches, such as cluster-model-based approaches, dateset-style methods, and ray tracing, face limitations in generalizability, spatial continuity, or computational efficiency~\cite{GSCM_eval,GSCM_joint_SAGE,hoydis2023sionna,alkhateeb2019deepmimo,zhang2024wisegrt}. In contrast, radio radiance field (RRF) methods have recently emerged as promising alternatives, as they capture the underlying physical principles of radio propagation rather than merely fitting observed outcomes~\cite{zhao2023nerf2,wen2024wrf,zhang2025rf3dgswirelesschannelmodeling}. For example, in our prior work, RF-3DGS~\cite{zhang2025rf3dgswirelesschannelmodeling}, we proposed a novel spatial wireless channel modeling framework that reconstructs the scene geometry from visual data and learns a continuous RRF using sparse wireless measurements. This enables accurate and efficient reconstruction of the Spatial-CSI, incorporating the path loss, delay, angle of arrival and departure (AoA and AoD) of strong paths, with fast training speed and real-time querying ability. RF-3DGS offers high spatial resolution, low memory requirements, and sub-second inference latency, making it highly suitable for real-time applications.

In this work, we further extend the RF-3DGS framework to address the unique challenges associated with modeling RRF in the THz band. Given that scattering NLoS paths are significantly attenuated at THz frequencies, accurately reconstructing the path loss becomes critical for enabling effective spatial beamforming. The proposed RF-3DGS+ framework estimates the full propagation path information based on scene geometry and leverages this to calibrate the path loss of each multipath component (MPC). 

Our main contributions are summarized as follows:
\begin{itemize}
\item We analyze the propagation characteristics of THz waves in complex environments and demonstrate how the RRF framework can effectively represent such spatial wireless channels. 
\item We propose RF-3DGS+, which addresses the ambiguity present in prior RRF-based methods by explicitly reconstructing full path information, including propagation distance and directional gains. 
\item To validate our approach, we develop a fine-grained THz simulation environment and conduct detailed experiments, demonstrating the superior accuracy and efficiency of RF-3DGS+ in modeling THz spatial channels.
\end{itemize}

\section{Related Work}
To model how radio waves propagate in 3D space, several categories of spatial channel modeling have been proposed.
Cluster-based models, such as those in~\cite{GSCM_eval,GSCM_joint_SAGE}, follow the geometry-based spatial-consistent MPC tracking approach. These models approximate radio wave propagation as rays under far-field assumptions and characterize key reflectors and scatterers in the environment, but they lack the fine-grained spatial resolution required for advanced 6G applications.
Ray tracing methods~\cite{hoydis2023sionna} also rely on geometric optics but use detailed environment reconstructions to simulate propagation. While these methods can yield accurate Spatial-CSI, their performance heavily depends on the quality of 3D meshes and surface electromagnetic (EM) property modeling. Furthermore, ray tracing is computationally intensive, limiting its real-time applicability.
Dataset-based methods~\cite{alkhateeb2019deepmimo,zhang2024wisegrt} store precomputed Spatial-CSI data at discrete 3D grid points, obtained from ray tracing or measurement campaigns. These methods enable fast querying and high fidelity, but they suffer from high data collection costs, limited generalizability across scenarios, significant memory requirements, and complex interpolation of  spatial attributes (e.g., AoA, AoD).

To understand Thz propagation characteristics, extensive measurement-based studies have been conducted~\cite{taleb2023scattering,sheikh2019scattering}. These works provide detailed physical insights into THz scattering behaviors, including scattering patterns and coefficients of different materials. However, such studies mainly serve as theoretical and simulation purposes, and their direct application to practical modeling remains limited.

Recently, neural radiance field (NeRF)-based methods have been introduced into the radio domain. NeRF is a popular approach for 3D reconstruction in computer vision, which describes the distribution of EM energy in the space. Since both optical and radio signals are EM waves, NeRF alike methods have started to be applied in wireless channel reconstruction (or modeling). NeRF$^2$\cite{zhao2023nerf2} was among the first to apply NeRF to wireless channels, jointly learning geometry and radio radiance via multi-layer perceptrons (MLPs), with Tx location conditioning. While effective on synthetic datasets, it suffers from high computational cost, slow inference, and the need for dense training data. NeWRF\cite{lu2024newrf} simplifies supervision by using a single complex-valued RSSI and estimated AoAs to infer virtual Tx positions and render radiance fields accordingly. Although innovative, its efficiency is still constrained by the NeRF backbone.

Following the introduction of 3D Gaussian Splatting (3DGS) in computer vision, several works have explored its adaptation to RF. WRF-GS~\cite{wen2024wrf} employs 3DGS to model wireless channels on the NeRF$^2$ dataset but lacks a dedicated geometry reconstruction stage, limiting accuracy. Scalable 3DGS~\cite{yang2025scalable} places Gaussians on predefined 3D grids, reducing learnable parameters. It replaces the spherical harmonics (SH) function with an MLP to model direction-dependent radiance, allowing for efficient rendering in high mobility scenarios such as vehicular communications.

\section{Modeling Terahertz Spatial Wireless Channel with RRF} 
\subsection{Characteristics of Terahertz Wireless Communication}
To analyze the spatial characteristics of THz wireless channels, we assume that the propagation between the Tx and Rx occurs under far-field conditions. This allows the use of geometric optics to approximate electromagnetic wave behavior, enabling the channel to be modeled as a multipath propagation model, where each MPC corresponds to a distinct propagation path (or ray) from the Tx to Rx. This multipath channel model can thus be expressed as:
\begin{equation} 
h(t) = \sum_{l=1}^{L} \alpha_l~e^{j\phi_l}~\delta(t - \tau_l), 
\label{eq1:multipath_channel} 
\end{equation}
\noindent where $\alpha_l$, $\phi_l$, and $\tau_l$ denote the amplitude, phase shift, and delay of the $l$-th MPC, respectively. 

At lower frequency, such as sub-6~GHz, specular reflection and diffraction dominate NLoS propagation, due to relatively smooth surfaces and sharp edges compared to the relatively long wavelengths. However, the situation changes drastically at the THz. As the wavelength is extremely short, surfaces appear relatively rougher and edges appear very dull. This severely reduces specular reflection and leads diffraction to be rare or negligible.
For example, as reported in~\cite{sheikh2019scattering}, at frequencies above 200~GHz, a plaster plane with a standard deviation in height of 0.25~mm and a correlation length of 1.8~mm exhibit near-zero specular reflection coefficients. In such cases, most of the reflected electromagnetic energy is scattered rather than preserved in the specular direction. As a result, scattering becomes the dominant mechanism for enabling NLoS coverage in the THz band. 


Another defining feature of THz communication is its extremely wide baseband. As specified in IEEE 802.15.3d-2017~\cite{serghiou2022terahertz}, the THz band between 252~GHz and 321~GHz is divided into channel bandwidths that are integer multiples of 2.16~GHz, ranging from $1\times$ to $32\times$ 2.16~GHz. This corresponds to sampling intervals ranging from 462.69~ps down to 14.47~ps, and multipath length resolution ranging from 13.9~cm to 0.4~cm. Under such fine temporal resolution, if the transmission is still omni-directional, the closely spaced scattering MPCs will result in distinct taps in the baseband channel impulse response (CIR). Given the ubiquity of THz scattering interactions, this leads to extremely long channel responses and significantly increases the overhead for channel estimation and equalization. In contrast, for lower-frequency systems such as LTE or Wi-Fi, which operate with basebands on the order of tens of MHz and time resolutions around tens of nanoseconds, such scattering paths tend to merge with stronger paths.

The above discussion highlights that, to ensure effective NLoS coverage in THz wireless systems, it is essential to exploit strong scattering paths while employing highly directional transmission and reception. This, in turn, demands fine-grained knowledge of the THz spatial wireless channel. 
For such purpose, directly modeling the scattering behaviour depends heavily on expensive and time-consuming measurement campaigns~\cite{taleb2023scattering,sheikh2019scattering}. More importantly, the resulting parameters are highly specific to surfaces with similar parameters, making them difficult to generalize or transfer. These limitations significantly hinder the practical deployment of such models in real-world scenarios. 
\begin{figure}[h]
    \centering
    \includegraphics[width=0.8\linewidth]{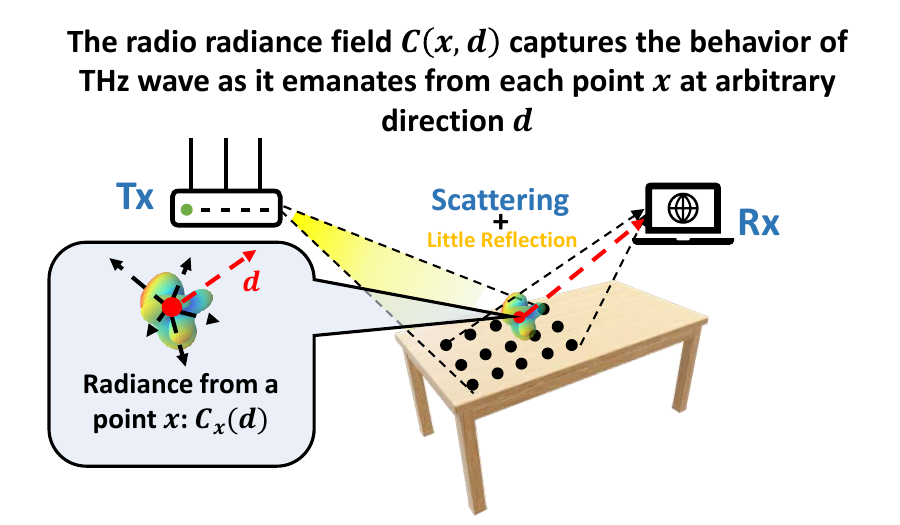}
    \caption{Modeling THz wave propagation with RRF.}
    \label{fig:enter-label}
\end{figure}

\subsection{Terahertz RRF Modeling and Reconstruction}

As an alternative to the traditional spatial channel models and the surface-specific measurements, the RRF offers an efficient and scalable solution for modeling THz spatial wireless channels in complex real-world environments.
As proposed in~\cite{zhang2025rf3dgswirelesschannelmodeling}, given a specific environment $\mathcal{E}$ and transmitter configuration $\mathcal{T}$, the RRF can be formulated as $\mathcal{R} =\{{\mathbf{c}(\mathbf{x_{obj}}, \mathbf{d}), \alpha(\mathbf{x})}\}$. Here, $\alpha(\mathbf{x})$ is a density field that encodes scene geometry: it assigns zero to free space, small values to translucent materials, and high values to solid surfaces. The function $\mathbf{c}(\mathbf{x_{obj}}, \mathbf{d})$ defines the Spatial-CSI of the radio radiance at each object point $\mathbf{x_{obj}}$ towards direction $\mathbf{d}$.


Returning to the multi-path channel model Eq.~\ref{eq1:multipath_channel}, each MPC corresponds to a radio path departing from the Tx with a specific AoD, interacting with $N$ intermediate object points $\mathbf{x_{reTx_i}}$ with corresponding outgoing directions $\mathbf{d_{i+1}}$, and finally being retransmitted from $\mathbf{x_{reTx_N}}$ toward the Rx along direction $\mathbf{d_{N+1}}$. This final retransmission ray arriving at the Rx can be denoted as $\mathbf{r}(t) = \mathbf{x_{reTx_N}} + t \mathbf{d_{N+1}}$. Obviously, the spatial-CSI value of this MPC can be represented by $\mathbf{c_{x_{reTx_N}, \mathbf{d_{N+1}}}}$, which includes the path loss, time-of-flight (ToF), AoD, and AoA of this MPC.

Noticeably, the same $\mathbf{c_{x_{reTx_N}, \mathbf{d_{N+1}}}}$ value is assigned to all MPCs that follow the same propagation path up to the final retransmitting point but terminate at different Rx positions along the same ray. These MPCs share identical directional properties and retransmission gains but differ in free-space path loss and ToF due to varying view depths $l_{vd}$, which are the distances from the Rxs to $\mathbf{x_{reTx_N}}$. While this discrepancy is negligible in visual radiance field domain, it becomes critical in the RF domain, especially in THz systems with ultra-wide bandwidth and high temporal resolution, where accurate ToF estimation is essential. This issue will be addressed in detail in later sections.

Then, for a Tx-Rx position pair, the multi-path channel model contains a large number of MPCs. To model such multi-path channel from the RRF $\mathcal{R}$, instead of integrating all incoming radiance over the full angular domain at the Rx, we adopt a more tractable approach by defining a finite set of AoAs to query—analogous to pixel directions in a camera model. For each rendering direction $\mathbf{d_{render}}$, the RRF integrates $\mathbf{c}(\mathbf{x_{obj}}, \mathbf{d_{render}})$ along the corresponding rendering ray via alpha-blending, which weights contributions based on the density field to ensure geometrically correct accumulation. The rendered Spatial-CSI values can then be organized into an Rx-side spatial spectrum, or an ``RF image", where each pixel direction corresponds to an AoA, and each color channel encodes path loss, ToF, and AoD.

During training, the rendered spatial spectrum is compared with the ground truth spectrum at corresponding Rx positions. The loss is then backpropagated through the differentiable alpha-blending process to update the learnable $\mathbf{c}(\mathbf{x_{obj}}, \mathbf{d_{render}})$ representation. To achieve efficient training and high-quality reconstruction, an optimized RRF representation and an efficient rendering pipeline are essential.

\section{Reconstructing Terahertz RRF with RF-3DGS+}
To address the challenges of modeling THz spatial wireless channels, we propose RF-3DGS+, which is tailored for capturing the THz scattering MPCs. The reconstruction is mainly based on a two-stage training pipeline: in the first stage, geometry is reconstructed from visual data; in the second stage, the RRF is further trained based on the training radio spatial spectra, and more importantly, the radio radiance rendering accounts for the full propagation path of each MPC, enabling unbiased estimation of both path loss and ToF.

\subsection{RF-3DGS+ Representation and Rendering}
In RF-3DGS+, the radiance field $\mathcal{R}=\{\mathbf{c}(\mathbf{x_{obj}}, \mathbf{d}), \alpha(\mathbf{x})\}$ is represented based on millions of 3D Gaussian primitives. Each Gaussian is parameterized by learnable geometric attributes: the base density $\alpha_g$, the center location $\mathbf{x}$, a scaling vector $\mathbf{s}_{scale}$, and a rotation quaternion $\mathbf{q}$.
The RRF function $\mathbf{c}(\mathbf{x_{obj}}, \mathbf{d})$ is defined per Gaussian, which approximates complex SH via weighted combinations of basis SH functions. Each Gaussian thus holds multiple SH coefficient sets, each set representing the angular distribution of a specific Spatial-CSI component. 

\begin{figure}
    \centering
    \includegraphics[width=0.75\linewidth]{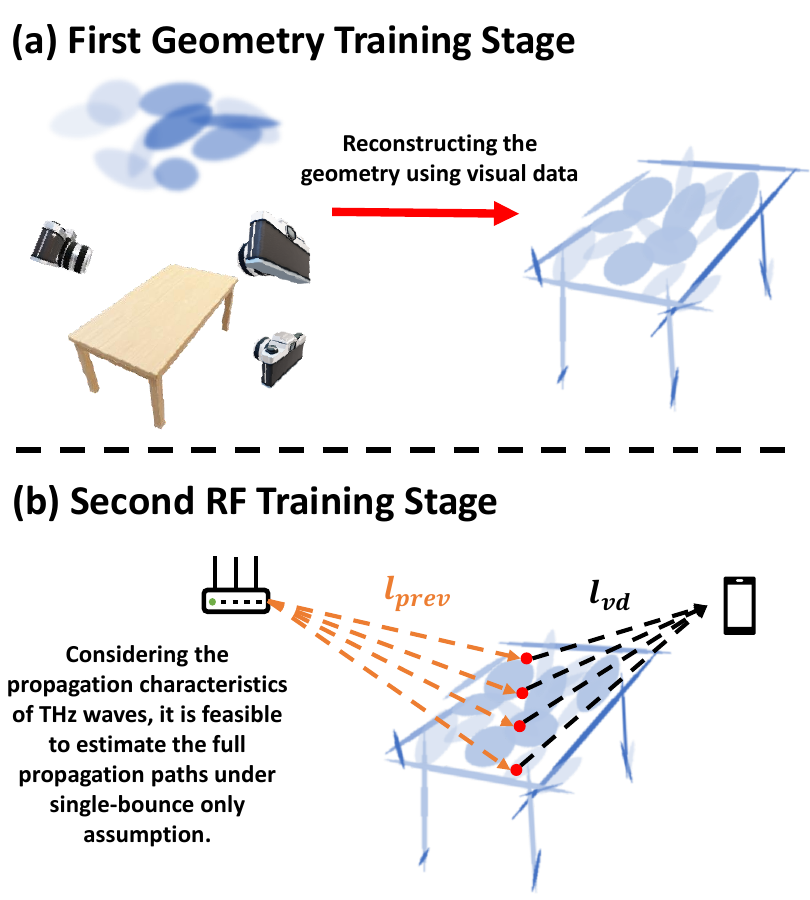}
    \caption{RF-3DGS+ pipeline.}
    \label{fig:enter-label}
    \vspace{-0.5cm}
\end{figure}

A critical limitation of the original RF-3DGS lies in its approximation: it shares the same Spatial-CSI across all Rx positions along the same outgoing ray, regardless of their actual view depth $l_{vd}$ to the final retransmitting point $\mathbf{x_{reTx_N}}$. This introduces notable deviations in path loss and ToF, especially in high-resolution THz scenarios. RF-3DGS+ resolves this by explicitly incorporating the full propagation path information into the radiance representation. Specifically, for each outgoing ray $\mathbf{r}(t) = \mathbf{x_{reTx_N}} + t \mathbf{d}_{N+1}$ emitted from a Gaussian, we record the cumulative prior path length $l_{prev}$ before reaching $\mathbf{x_{reTx_N}}$, the AoD at the Tx,  and the cumulative interaction gain $\prod_{i=1}^{N-1} A_i$ up to the last bounce.
Accordingly, the SH function at each Gaussian now only models the final directional interaction gain $A_N(\mathbf{d_{render}})$. During rendering, the total propagation distance is obtained by summing $l_{prev}$ and the current view depth $l_{dp}$, enabling unbiased free-space path loss and ToF computation.

To render the Spatial-CSI at a receiver location $\mathbf{x_{Rx}}$, we define a discrete set of AoA directions. Instead of performing rendering on a per-ray basis, the Gaussians are splatted onto the rendering plane in the order of their view depth, enabling parallel rendering across all affected pixels. However, when consider the rendering along a specific rendering direction $\mathbf{d_{render}}$, the received Spatial-CSI is computed via alpha-blending of the $K$ visible Gaussians along the ray $\mathbf{r_{render}}(t) = \mathbf{x_{Rx}} + t\mathbf{d_{render}}$. Notably, in RF-3DGS+, since the full propagation path information is explicitly recorded, the rendering process only needs to compute the path loss $ \mathbf{p}_{recv} $ for each direction, and this can be expressed as:
\begin{align}
    {p_{recv}}(\mathbf{x_{Rx}}, \mathbf{d_{render}}) 
    &= \sum_{k=1}^{K} \alpha_k \cdot T_k  \cdot p_k(\mathbf{d_{render}}), \label{eq:alpha-blending} \\[6pt]
    T_k 
    &= \exp\left(-\sum_{j=1}^{k-1} \alpha_j\right), \label{eq:transparency} \\[6pt]
    p_k(\mathbf{d_{render}}) 
    &= \left( \prod_{i=1}^{N-1} A_i \right) \cdot A_N(\mathbf{d_{render}}) \notag \\
    &\quad \cdot \left( \frac{\lambda}{4\pi(l_{prev}+l_{vd})} \right)^2~. \label{eq:pathloss}
\end{align}
Here, $\alpha_k$ denotes the effective density of the $k$-th Gaussian along the ray, which is computed as the product of its base density $\alpha_g$ and the approximated density of its intersection with the rendering ray. The term $T_k$ represents the accumulated transmittance, accounting for the occlusion caused by all preceding Gaussians along the ray. The path loss contribution $p_k(\mathbf{d}_{\text{render}})$ from the $k$-th Gaussian is computed according to Eq.~\ref{eq:pathloss}, where the denotations are simplified but both the interaction gains and the propagation lengths are specific to that particular Gaussian. It is also assumed that the training RF data has been preprocessed to normalize the directional antenna gain.

\subsection{Reconstruction of RRF and Full Path Information}
Given the representation and rendering mechanism of RF-3DGS+, this section introduces how RF-3DGS+ reconstructs the geometry, radio radiance, and full path information specifically for THz communication scenarios.

Accurate geometry reconstruction from RF data alone is highly challenging. Therefore, RF-3DGS+ adopts a two-stage fusion training strategy. 
In the first stage, RF-3DGS+ performs visual-based geometry reconstruction through RRF training. Visual images are rendered from the current learnable representation at known poses, and the corresponding loss is computed with respect to the corresponding ground-truth images. This loss is then backpropagated through the differentiable rendering pipeline. During this process, the 3D Gaussians are dynamically updated—translated, rotated, split, or pruned—to fit the observed geometry. After convergence, the geometry-related parameters are frozen for the subsequent RF radiance training. In the second stage, the goal is to reconstruct full path information, which includes  the prior path length $l_{prev}$, the AoD, the interaction gain $\prod_{i=1}^{N-1} A_i$, and the final directional gain $A_N(\mathbf{d_{render}})$. This reconstruction is particularly difficult, even when geometry is known. In the original RF-3DGS framework, delay and AoD spectra are required as training input to reconstruct these components, which incurs significant data acquisition costs.

However, thanks to the nature of THz wave propagation, this process can be greatly simplified. At THz frequencies, scattering becomes the dominant propagation mechanism in NLoS scenarios, while multi-bounce paths with two or more scattering reflections are usually negligible due to their high cumulative path loss. Under this assumption, we only need to consider single-bounce paths, which significantly reduces complexity. In such case, there is only one interaction, and $l_{prev}$ can be easily computed as the distance between the intersection point on the Gaussian ray and the Tx position, which is typically known from the RF measurement setup. To further improve efficiency, we approximate the interaction point on each ray using the center of the Gaussian with the highest contribution (i.e., the product of the transmittance and effective density) along that ray. This point is treated as the pseudo-surface point and used to compute $l_{prev}$, the view depth $l_{vd}$, and the AoD for all Gaussians along that ray. Since the Gaussians have limited spatial extent (with abnormally large ones pruned during first stage), this approximation introduces only minor errors while significantly accelerating computation. Then, during second RF training stage, the rendered path loss spectra are compared against the ground truth and used as training supervision. The SH functions of each Gaussian are optimized to fit the directional gain component $A_N(\mathbf{d_{render}})$.

\section{Simulation and Experimental Results}
To evaluate the effectiveness of RF-3DGS+ in THz communication scenarios, we construct a comprehensive simulation based on the Sionna wireless simulator developed by NVIDIA~\cite{hoydis2023sionna}, which supports GPU-accelerated ray tracing with meticulous EM parameters. Specifically, we build a high-fidelity 3D mesh model of an indoor lobby environment featuring complex geometries. For each surface, we carefully configure the EM parameters such as relative permittivity, conductivity, and surface roughness related parameters, such as scattering coefficients and pattern. These parameters are derived from existing THz measurement studies~\cite{taleb2023scattering,sheikh2019scattering}, ensuring consistency with empirical data.

We fix the Tx position and uniformly sample 800 Rx positions throughout the environment to construct a spatially diverse dataset. From this pool, 100 Rx positions are randomly selected as a held-out test set, which remains consistent across all experiments. The remaining 700 samples are used to create training subsets of varying sizes: 10, 15, 20, 25, 50, 100, 200, 400, and 700 samples. This allows us to assess model performance under different levels of data sparsity, which is critical in practical deployment scenarios.

All experiments are conducted on a desktop equipped with an AMD Ryzen 7700 CPU and an NVIDIA RTX 4090 GPU.
To benchmark the performance of RF-3DGS+, we compare it against three representative baseline methods. The first is the original RF-3DGS, which does not incorporate full path information and serves as a geometry-aware baseline. The second is NeRF$^2$, a NeRF-based approach adapted for RF applications, representing the broader class of neural rendering techniques. The third baseline is a conditional GAN (CGAN), which serves as a representative data-driven generative model that learns the mapping from Rx positions to spatial spectra without explicit geometry modeling.

\begin{figure}[h]
    \vspace{-0.3cm}
    \centering
    \includegraphics[width=0.8\linewidth]{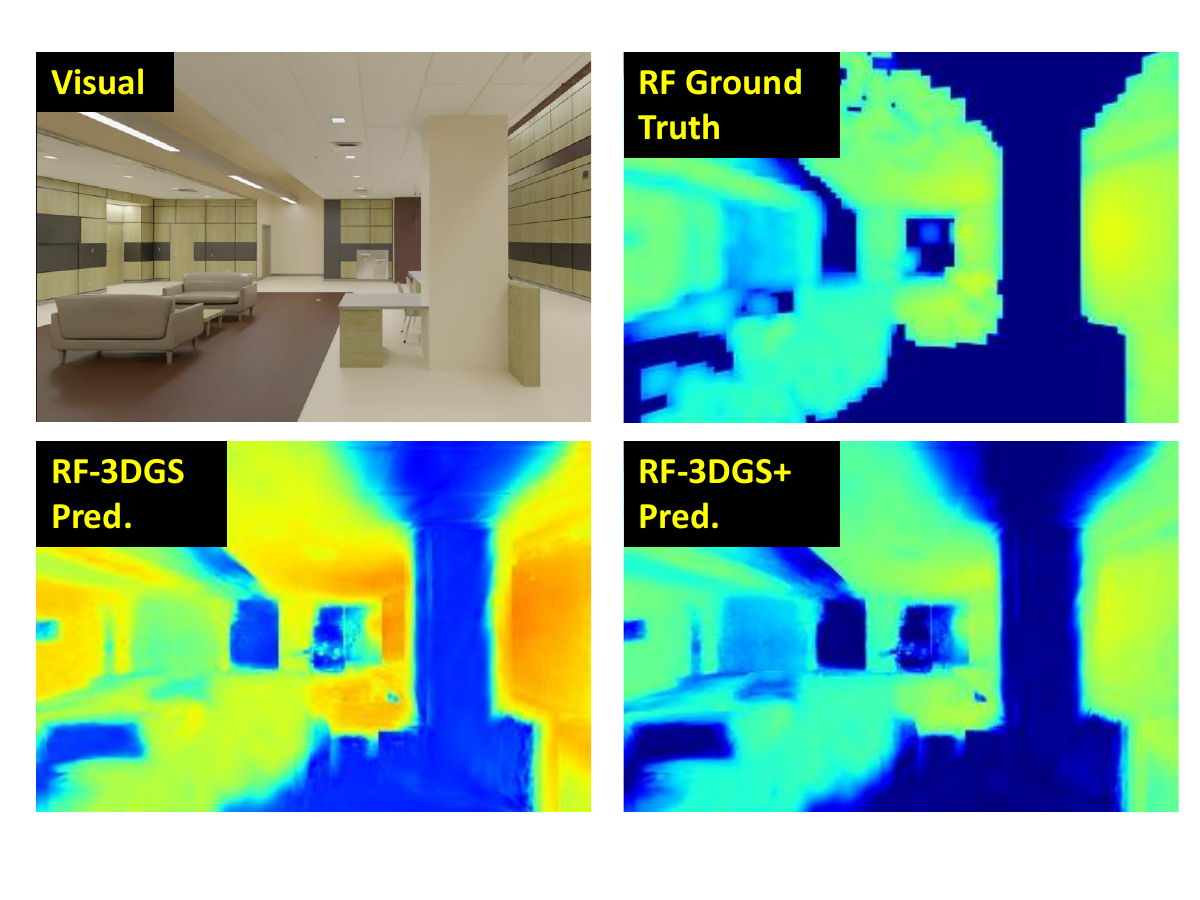}
    \vspace{-0.7cm}
    \caption{A visual illustration of THz reconstruction}
    \vspace{-0.2cm}
    \label{fig:beamforming}
\end{figure}
Table~\ref{tab:general_comparison} reports the average performance of all methods across several key metrics that are popular in computer vision, including peak signal-to-noise ratio (PSNR, the higher the better), structural similarity index measure (SSIM, the higher the better), and learned perceptual image patch similarity (LPIPS, the lower the better), to assess reconstruction fidelity. Additionally, we evaluate the training time and inference speed of each method to analyze their computational efficiency and suitability for real-time or large-scale deployment.
\begin{table}[h]
\caption{General comparison between RF-3DGS+ and baseline methods.}
\label{tab:general_comparison}
\centering
\begin{tabular}{lccccc}
\hline
\textbf{Method} & \textbf{PSNR} $\uparrow$ & \textbf{SSIM} $\uparrow$ & \textbf{LPIPS} $\downarrow$ & \textbf{Train} $\downarrow$ & \textbf{Infer.} $\downarrow$ \\
\hline
    RF-3DGS+   & \textbf{19.68} & \textbf{0.635} & \textbf{0.439} & 3m~42s & 3.4~ms \\
    RF-3DGS    & 13.50 & 0.476 & 0.456 & \textbf{2m~47s} & \textbf{2~ms} \\
    NeRF$^2$   & 13.34 & 0.463 & 0.691 & 2h~32m & 0.7~s \\
    CGAN       & 10.33 & 0.212 & 0.874 & 1h~27m & 5~ms \\
\hline
\end{tabular}
\end{table}

As shown in the results, RF-3DGS+ outperforms the baseline methods across evaluation metrics. It achieves much higher reconstruction quality, indicated by superior PSNR, SSIM, and LPIPS scores, while also demonstrating significantly shorter training and inference times compared to NeRF$^2$ and CGAN. Thus, in the remaining part, we mainly focus on detailed comparisons between RF-3DGS+ and RF-3DGS to further analyze the specific improvements introduced by our proposed enhancements.
\begin{figure}[h]
    \centering
    \includegraphics[width=0.7\linewidth]{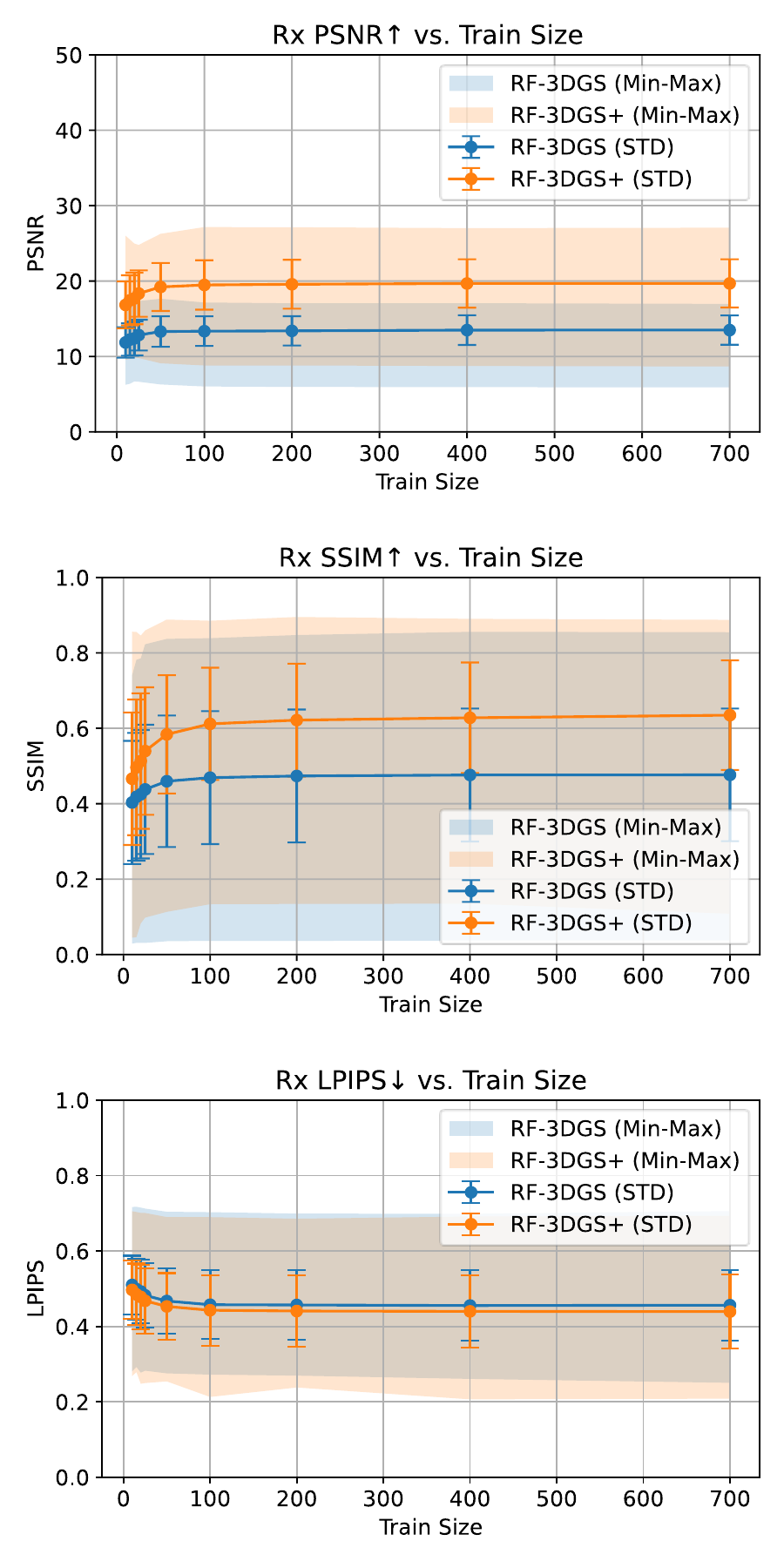}
    \vspace{-0.3cm}
    \caption{Performance comparison with existing works}
    \vspace{-0.3cm}
    \label{fig:detail metric}
\end{figure}

As shown in Fig.~\ref{fig:detail metric}, we compare the reconstruction performance of RF-3DGS+ and the original RF-3DGS across various training set sizes. The results clearly demonstrate that RF-3DGS+ consistently outperforms RF-3DGS in terms of both PSNR and SSIM across all training sizes. Not only are the mean values higher, but the minimum-maximum ranges are also consistently higher. This validates the effectiveness of our full-path-aware modeling approach for THz spatial channel reconstruction. The observed performance gain is primarily attributed to RF-3DGS+'s ability to recover the full propagation path information which is especially critical for THz communication due to its ultra-high temporal resolution. These results also empirically support our earlier assumption that in typical indoor THz environments, single-scattering paths dominate NLoS propagation. Moreover, RF-3DGS+ exhibits strong resilience to training data sparsity, maintaining competitive performance even when trained with as few as 20 samples. Only when the training size drops below this threshold does performance degrade significantly, indicating a minimal data requirement for reliable generalization.

On the other hand, the improvement in LPIPS is relatively modest. This is expected, as LPIPS is a perceptual similarity metric that leverages deep neural network features to compare high-level visual characteristics, emulating human visual perception. In contrast to PSNR and SSIM, which mainly evaluate pixel-level and statistical structural consistency, LPIPS is more sensitive to discrepancies in geometry alignment. In our case, both RF-3DGS and RF-3DGS+ achieve comparable geometry reconstruction quality due to the shared visual-based geometry initialization in the first training stage. The main difference lies in RF-3DGS’s inability to capture the path loss variation caused by different view depths, a limitation that affects PSNR and SSIM but has less impact on LPIPS. 


\section{Conclusion}
In this work, we proposed RF-3DGS+, an RRF framework tailored for THz communication scenarios where single-bounce scattering paths dominate. By incorporating full path length into the rendering process, RF-3DGS+ significantly improves reconstruction accuracy with minimal training data. The method addresses key limitations of previous 3D Gaussian-based models, especially the approximation of view depth. While currently limited to single-bounce paths, RF-3DGS+ is well-suited for short-range and indoor THz applications, and lays the groundwork for future extensions to multi-bounce scenarios and improved geometric representations.

\bibliographystyle{IEEEtran}
\bibliography{references}
\end{document}